\documentclass[aps,pre,twocolumn,superscriptaddress,floats,floatfix,showpacs]{revtex4-1}
\usepackage{amssymb,amsmath}
\usepackage{float}
\usepackage{mathrsfs}
\usepackage{graphics}
\usepackage{epstopdf}
\usepackage{color}
\usepackage{subfigure}
\usepackage{epsfig}
\usepackage{rotating}
\usepackage{anyfontsize}
\usepackage{verbatim}
\usepackage{standalone}
\usepackage{hyperref}

\usepackage{xr}
\begin{document}

\title{An {\it in silico} study of electrophysiological parameters that affect the spiral-wave frequency in mathematical models for cardiac tissue}

\author{Mahesh Kumar Mulimani}
\email{maheshk@iisc.ac.in ;}
\affiliation{Centre for Condensed Matter Theory, Department of Physics, Indian Institute of Science, Bangalore 560012, India.}
\author{Soling Zimik}
\email{solyzk@gmail.com ;}
\affiliation{Centre for Condensed Matter Theory, Department of Physics, Indian Institute of Science, Bangalore 560012, India.}
\author{Rahul Pandit}
\email{rahul@iisc.ac.in}
\altaffiliation[\\]{also at Jawaharlal Nehru Centre For
Advanced Scientific Research, Jakkur, Bangalore, India}
\affiliation{Centre for Condensed Matter Theory, Department of Physics, Indian Institute of Science, Bangalore 560012, India.}

 
\begin{abstract}

Spiral waves of excitation in cardiac tissue are associated with
life-threatening cardiac arrhythmias. It is, therefore, important to
study the electrophysiological factors that affect the dynamics of
these spiral waves. By using an electrophysiologically detailed
mathematical model of a myocyte (cardiac cell), we study the effects of
cellular parameters, such as membrane-ion-channel conductances, on the
properties of the action-potential (AP) of a myocyte. We then
investigate how changes in these properties, specifically the upstroke
velocity and the AP duration (APD), affect the frequency $\omega$ of a
spiral wave in the mathematical model that we use for human-ventricular
tissue. We find that an increase (decrease) in this upstroke-velocity
or a decrease (increase) in the AP duration increases (decreases)
$\omega$. We also study how other intercellular factors, such as the
fibroblast-myocyte coupling, diffusive coupling strength, and the effective
number of neighboring myocytes, modulate $\omega$. Finally,
we demonstrate how a spiral wave can drift to a region with a high
density of fibroblasts.  Our results provide a natural explanation for
the anchoring of spiral waves in highly fibrotic regions in fibrotic
hearts.    
\end{abstract}

\maketitle

\section{Introduction}
\label{sec:intro}

Nonlinear waves in the form of rotating spirals are ubiquitous spatiotemporal
patterns that occur in a variety of biological or physical systems; these
include chemical-reaction waves in the Belousov-Zhabotansky system
~\cite{zaikin1970concentration,winfree1972spiral,field1993chaos,ott2002chaos,strogatz2018nonlinear},
oxidation waves of carbon monoxide on the surface of
platinum~\cite{falcke1992traveling,imbihl1995oscillatory,pande1999spatiotemporal},
calcium-signalling waves in Xenopus oocytes~\cite{lechleiter1991spiral},
cyclic-AMP signalling waves in the aggregration process of Dictyostelium
discoideum~\cite{tyson1989cyclic,rietdorf1996analysis}, and, notably,
action-potential (AP) waves that mediate muscle contraction in cardiac tissue.
The organization of these AP waves in the form of spirals or scrolls in cardiac
tissue is associated with abnormal and life-threatening heart rhythms known as
arrhythmias.  In particular, ventricular arrhythmias can lead to sudden cardiac
death; therefore, it is important to understand the dynamics of such waves.    

The rhythm of a normal heart is maintained by the trains of waves that are
generated by its pacemaker, the sino-atrial node (SAN). This normal rhythm in a
heart can be disturbed by the formation of a spiral wave, which can override
the function of the SAN as the primary source of waves and entrain the heart to
follow the spiral-rotation frequency. There are multiple mechanisms through
which spiral waves can occur in cardiac
tissue~\cite{panfilov1991vortex,lim2006spiral,xie2009effects,qu2014nonlinear,nayak2013spiral,nayak2015turbulent,zimik2017reentry}.
A single-spiral state is associated with ventricular tachycardia (VT), which
leads to a fast heart rate. A multiple-spiral state is linked to ventricular
fibrillation (VF) that results in a chaotic heart
rate~\cite{davidenko1992stationary,ten2003reentry,lim2006spiral,xie2009effects,clayton2011models,qu2014nonlinear,alonso2016reentry},
and a quivering of the left ventricle, which renders it incapable of pumping
oxygenated blood to the body; and, in the absence of medical intervention, this
leads to death in a few minutes. It is crucial, therefore, to develop a
detailed understanding of how spiral waves in cardiac tissue can get
destabilized and form multiple spiral waves. Some studies have shown that
heterogeneity-induced spatial gradients in the frequency $\omega$ of a spiral
wave can lead to such an
instability~\cite{zimik2016instability,zimik2020anisotropic} or to the drifting
of this spiral
wave~\cite{krinsky1996dense,ten2003reentry,sridhar2010anomalous,biktashev2011evolution}.
We build on the results of these studies to investigate which physiological
factors affect $\omega$ and how they modulate it. In mammalian hearts, cardiac
tissue is heterogeneous: there can be cellular heterogeneity, e.g., cardiac
fibroblasts in addition to myocytes, or a spatial variation of
electrophysiological properties, e.g., along the apico-basal direction in a
heart, or between intermural layers~\cite{wolk1999functional} of the heart, or
because of conduction inhomogeneities~\cite{ten2005wave}.

We investigate the effects of various intracellular (ion-channel conductances)
and intercellular (gap-junctional factors) parameters on the spiral-wave
frequency.  At the single-cell level, we show how changes in ion-channel
conductances modulate action-potential (AP) properties, such as its upstroke
velocity $\frac{dV}{dt}_{max}$ and duration (APD). We then examine how these
changes in AP properties affect the spiral-wave frequency $\omega$ at the
tissue level.  We find that an increase (decrease) in $\frac{dV}{dt}_{max}$
(APD) increases (decreases) $\omega$. We then investigate the effects of
intercellular coupling strength on $\omega$ by changing the coupling strength
in the following two ways: (a) by modifying the diffusion constant $D$ of the
medium; (b) by interspersing inexcitable point obstacles in the medium, thereby
reducing the effective number of neighboring myocytes. We find that  $\omega$
is unaffected by a change in $D$, but, with point obstacles, $\omega$ decreases
with an increase in the density of these obstacles. We examine two models for
fibrosis, which occurs in diseased hearts and is usually accompanied by a
proliferation of
fibroblasts~\cite{weber1994collagen,manabe2002gene,gurtner2008wound,biernacka2011aging,hinderer2019cardiac}.
These models allow us to study how various fibroblast parameters, e.g., the
fibroblast-myocyte coupling and the AP of the coupled myocyte, affect $\omega$
and spiral-wave dynamics; the fibroblast parameters include its resting
potential and the number of fibroblasts coupled to a myocyte.  Moreover, we
show that a spiral in a medium with a heterogeneous distribution of
fibroblasts, drifts towards the region with a high density of fibrolasts.

The paper is organized as follows. The Materials and Methods
Section~\ref{sec:materials} contains (a) the details of the myocyte and tissue
models that we use in our simulations and (b) the numerical techniques we use
to solve the governing equations. We then provide the findings of our study in
the Section~\ref{sec:results} on Results. Finally, in the Discussion
Section~\ref{sec:discussion}, we discuss our results in the light of other past
studies and mention some of the limitations in our study.

\section{Materials and Methods}
\label{sec:materials}

For myocytes we use the TP06 human-ventricular-cell
model~\cite{ten2006alternans}, in which the transmembrane potential $V_{m}$ of
an isolated myocyte is governed by the following ordinary differential equation
(ODE):
\begin{eqnarray}	 
\frac{dV_{m}}{dt} &=& -\frac{I_{ion}}{C_{m}} \label{eq:Vm}; \nonumber \\
I_{ion} &=& \sum_{i} I_{i};
\label{Model:myocyte}
\end{eqnarray}
$I_{ion}$ is the sum of all the ion-channel currents with $I_{i}$ the $i^{th}$
ion-channel current, and $C_m$ the normalized transmembrane capacitance. 
In Table~\ref{Table1} we list the currents in the TP06 model; their dependence
on $V_m$ is given, e.g., in Ref.~\cite{ten2006alternans}.

\begin{table}
\begin{center}
    \begin{tabular}{ | l |  p{6.5cm} |}
    \hline 
    $I_{Na}$ & fast inward $Na^+$ current  \\ \hline 
    $I_{CaL}$ & L-type inward $Ca^{++}$ current   \\ \hline
    $I_{to}$ & Transient outward current   \\ \hline
    $I_{Ks}$ & Slow delayed rectifier outward $K^+$ current    \\ \hline
    $I_{Kr}$ & Rapid delayed rectifier outward $K^+$ current   \\ \hline 
    $I_{K1}$ & Inward rectifier outward $K^+$ current   \\ \hline
    $I_{NaCa}$ & $Na^+ / Ca^{++}$ exchanger  current  \\ \hline
    $I_{NaK}$ & $Na^+ /K^+$ pump current    \\ \hline
    $I_{pCa}$ & plateau $Ca^{++}$ current  \\ \hline 
    $I_{pK}$ & plateau $K^+$ current   \\ \hline
    $I_{bNa}$ & background inward $Na^+$ current   \\ \hline
    $I_{bCa}$ & background inward $Ca^+$ current    \\ \hline
    
    \end{tabular}

 \hspace{7cm}\caption{\textbf{The various ionic currents in the TP06 
model~\cite{ten2006alternans}; the symbols used for the currents follow 
Ref.~\cite{ten2006alternans}. }}
\end{center}
\label{Table1}
\end{table}
The spatiotemporal evolution of $V_m$ in mathematical models for cardiac tissue 
is governed by the following reaction-diffusion partial differential equation (PDE):
\begin{equation}
\frac{\partial{V_{m}}}{\partial{t}} = D\ {\nabla^2}V - \frac{I_{ion}}{C_{m}} ;
\label{eq:VmPDE}
\end{equation}
$D$ is the diffusion coefficient; we restrict ourselves to a scalar $D$ for simplicity;
the TP06 case is described in detail in their dependence
on $V_m$ is given, e.g., in Refs.~\cite{ten2006alternans}.
%%%%%%%%%%%%%%%%%%%%%%%%%%%%%%%%%%%%%%%%%%%%%%%%%%%%%%%%%%%%%%%%%%%%%%%%%%%%%%
It is convenient to use the following non-dimensionalised
ion-channel conductances and diffusion coefficients:
\begin{equation}
S_G = \frac{G}{G_{c}} \; ; S_D = \frac{D}{D_0} , 
\label{eq:nondim}
\end{equation}
where $G$ stands for a typical conductance, $G_{c}$ is the control value
of the conductance, and the control diffusion constant $D_0=0.00154 \ cm^2/ms$;
for the conductances we consider, $G_{c} = 14.838, \, 0.0000398, \, 0.153$
pA/pF for $G_{Na}, G_{CaL},$ and  $G_{Kr}$, respectively.

We use the following two models for the fibroblast cells:
\begin{itemize}

\item \textbf{Model I}: We model the fibroblast cells as inexcitable obstacles
and we replace the myocytes at random with these inexcitable obstacle
throughout our simulation domian such that the percentage of
sites with obstacles is $p_o$. The gap-junctional
current between the myocyte and and 
inexcitable obstacles in this model is zero (see Refs.~\cite{ten2005wave,majumder2012nonequilibrium}).

\item \textbf{Model II}:We model the fibroblasts in our study as an
electrically passive cell, as in Ref.~\cite{maccannell2007mathematical}.
Each myocyte is coupled to $N_f$ fibroblasts; and
the myocyte and fibroblast transmembrane potentials $V_m$ and $V_f$,
respectively, obey the following coupled ODEs:
\begin{eqnarray}
\frac{dV_{m}}{dt} &=& -\Big{(}\frac{I_{ion}}{C_{m}}+N_f \times \frac{I_{gap}}{C_{m}}\Big{)}; \label{eq:Vf}  \nonumber \\
\frac{dV_{f}}{dt} &=& \frac{(I_{gap}-I_{f})}{C_{f}};  \nonumber \\
I_{f} &=& G_f \ (V_f-E_f); \nonumber \\
I_{gap} &=& G_{gap} \ (V_m-V_f).
\label{eq:VmVf}
\end{eqnarray}

$C_f$, $E_f$, and $G_{gap}$ are the membrane capacitance of a fibroblast,
the fibroblast resting potential, and the fibroblast-myocyte gap-junctional 
coupling, respectively. We use a \textit{bilayer model} for 
fibroblast-myocyte couplings: fibroblasts, in the top layer, are coupled to myocytes 
in the bottom layer, as in Ref.~\cite{nayak2013spiral}, which contains a schematic 
diagram of this bilayer and the PDEs that describe
the spatiotemporal evolution of waves of activation in this model; we do not include 
fibroblast-fibroblast couplings. Moreover, when we consider a heterogeneous 
distribution of fibroblasts in Sec.~\ref{subsec:drift}, we remove randomly all
the $N_f$ fibroblasts at a site in the top layer, so that the percentage of i
sites 
at which we retain fiboblasts is $p_f$. To study gradients in the density 
of fibroblasts, we use a space-dependent density that varies linearly
as we move away from chosen central site:
\begin{equation}
p_f(r_i)= p_f(r_0)- \frac{[p_f(r_0)-p_f(r_{max})]}{[r_{max}-r_{0}]} \times r_i ,
\label{eqn:eqn1}
\end{equation}
where $r_i$ is the distance from the centre, $r_0$ is the position of the
centre, and $r_{max}$ is maximum radial distance from the
centre.  
\end{itemize}

We update the ODEs via the forward-Euler method for Eqs.~\ref{eq:Vm} and
~\ref{eq:VmVf}.  For our two-dimensional (2D) tissue simulations as in
Eq.~\ref{eq:VmPDE} we use a square domain with $ N \times N$ grid points with
$N = 512$, the forward-Euler scheme for time marching, and a central-difference
scheme with a five-point stencil for the Laplacian, with the time and space
steps $\Delta t = 0.02 \ ms$ and $\Delta x = 0.025 \ cm$, respectively. The
control value of the diffusion coefficient $D=D_0$, in Eqs.~\ref{eq:VmPDE}, is
$D_0=0.00154 \ cm^2/ms$, which gives us a conduction velocity $CV \simeq 70 \
cm/s$, as has been reported for human-ventricular-tissue
models~\cite{ten2004model,ten2006alternans}.

\begin{itemize}
\item We calculate the frequency $\omega$ by recording the time-series of the
transmembrane potential $V_m$ at four representative positions in the
simulation domain. From the principal peak in the Fourier
transforms of these time series, we obtain $\omega$ (we take
the average of the values at the four representative positions). [We show in 
Table S1  of the Supplemental Material~\cite{supmat_EP} that this 
frequency is within error bars of the frequency 
$\omega_{tip}$ of rotation of the tip of the spiral wave.]
\item For the radius of the tip trajectory of a rigidly rotating spiral waves,
which is, on average, circular, we fit the average trajectory to a
circle with radius $r$ and center $(x_c,y_c)$, by using a nonlinear
regression model, to obtain the mean radius and the mean values of the
coordinates of the center of the circle; we also calculate the standard
deviation of the fluctuations in $r$ by using the mean position of the
center $(x_c,y_c)$ and the coordinates $(x,y)$ of the points that lie
on the unaveraged tip trajectory that we compute .
\item We calculate CV by pacing the simulations domain at one end with a pacing
cycle length of $1$ Hz; we use $20$ pulses. We record the time series
of $V_m$ at two designated grid points $A$ and $B$, which are
separated by a distance $l_{AB}$. These grid points are chosen
such that the line between the two grid points is normal to the
wavefront. We obtain the times $t_A$ and $t_B$ at which the
wavefront hits the grid points $A$ and $B$, respectively; the
difference $t_B-t_A$ gives the time taken by the wavefront to
propagate $A$ to $B$; therefore, CV $= \frac{l_{AB}}{(t_B-t_A)}$.
In the disordered case, with inexcitable obstacles distributed
at random in the simulation domain, we record the time series
of $V_m$ at multiple points and repeat the above procedure; we
then take the mean of the CVs obtained from these points; we
also compute the standard deviation of the CVs.
\end{itemize}

\section{Results}
\label{sec:results}

We present the results of our \textit{in-silico} studies as follows: In
Subsection~\ref{subsec:AP} we examine the dependence of the AP and of $\omega$
on various ion-channel conductances. Subsection~\ref{subsec:gapjn} is devoted
to the effects of the gap-junctional coupling on $\omega$. In
Subsection~\ref{subsec:sec3} we investigate the effects of the
fibroblast-myocyte coupling on the myocyte AP and $\omega$. We elucidate the
drift of spiral waves in domains with an inhomogeneous distribution of
fibroblasts in Subsection~\ref{subsec:drift}.

\subsection{Effects of conductances on the AP and the spiral-wave frequency
$\omega$}
\label{subsec:AP}

\begin{figure*}
  \centering
	\includegraphics[scale=0.18]{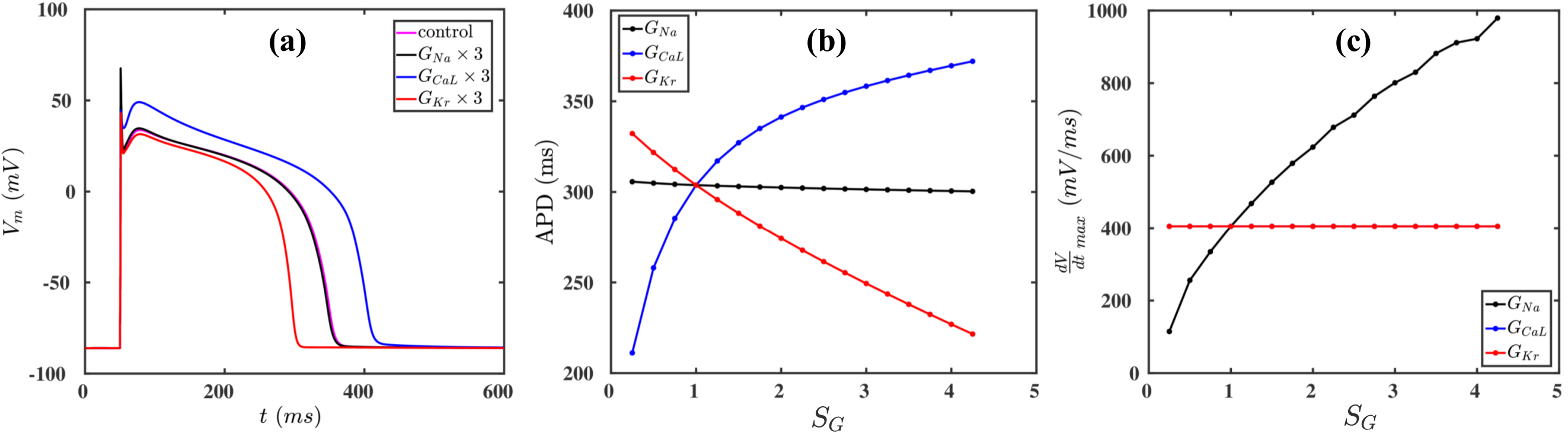}

\caption{(a) Action-potential plots for the control paramater set (magenta) and
the cases when the conductances $G_{Na}$ (black),  $G_{CaL}$ (blue),
and $G_{Kr}$ (red) are increased by a factor of three relative
to their control values. (b) and (c): Plots of the APD and $\frac{dV}{dt}_{max}$, 
respectively, versus $S_G$, the non-dimensionalized conductance~\ref{eq:nondim}.}	
\label{fig:fig1}	
\end{figure*}

The cell membrane of a myocyte is embedded with various ion-channels, which we
list in Table~\ref{Table1}; $V_m$ depends on the currents through these
ion-channels (Eq.~\ref{eq:Vm}), so, if we vary the conductances of these
channels, we can modulate the AP of the myocyte. To study the effects of these
ion channels on the AP, we choose three representative major ionic currents for
our study: $I_{Na}$, $I_{CaL}$, and $I_{kr}$. Figure~\ref{fig:fig1} (a) shows
the APs of a myocyte for control values (magenta) and for the cases where the
conductances $G_{Na}$ (black), $G_{CaL}$ (blue), and $G_{Kr}$ (red) are
increased three-fold. We find that increasing $G_{CaL}$ ($G_{Kr}$) increases
(decreases) the APD, whereas $G_{Na}$ has no significant effect on the APD
(Fig.~\ref{fig:fig1} (b)). This is because the inward current $I_{CaL}$
augments depolarization, and $I_{Kr}$, being an outward current, enhances
repolarization; although $I_{Na}$ is an inward current, it is active only
during the early upstroke phase of the AP, therefore, it cannot affect the APD
siginificantly. Futhermore, we find that increasing $G_{Na}$ increases the
upstroke velocity $\frac{dV}{dt}_{max}$, but $G_{CaL}$ and $G_{Kr}$ do not
affect on $\frac{dV}{dt}_{max}$ (see Fig.~\ref{fig:fig1}). We have also checked
the effects of other ion-channel conductances and ion-pump parameters on the
AP. The results are consistent with our findings above, namely, increasing
(decreasing) the conductances of inward (outward) currents increases
(decreases) the APD of the myocyte; and $I_{Na}$ is the only current that can
change the value of $\frac{dV}{dt}_{max}$. We give details in Fig. S1 in the
Supplemental Material~\cite{supmat_EP}. 

\begin{figure*}
  \centering	
	\includegraphics[scale=0.25]{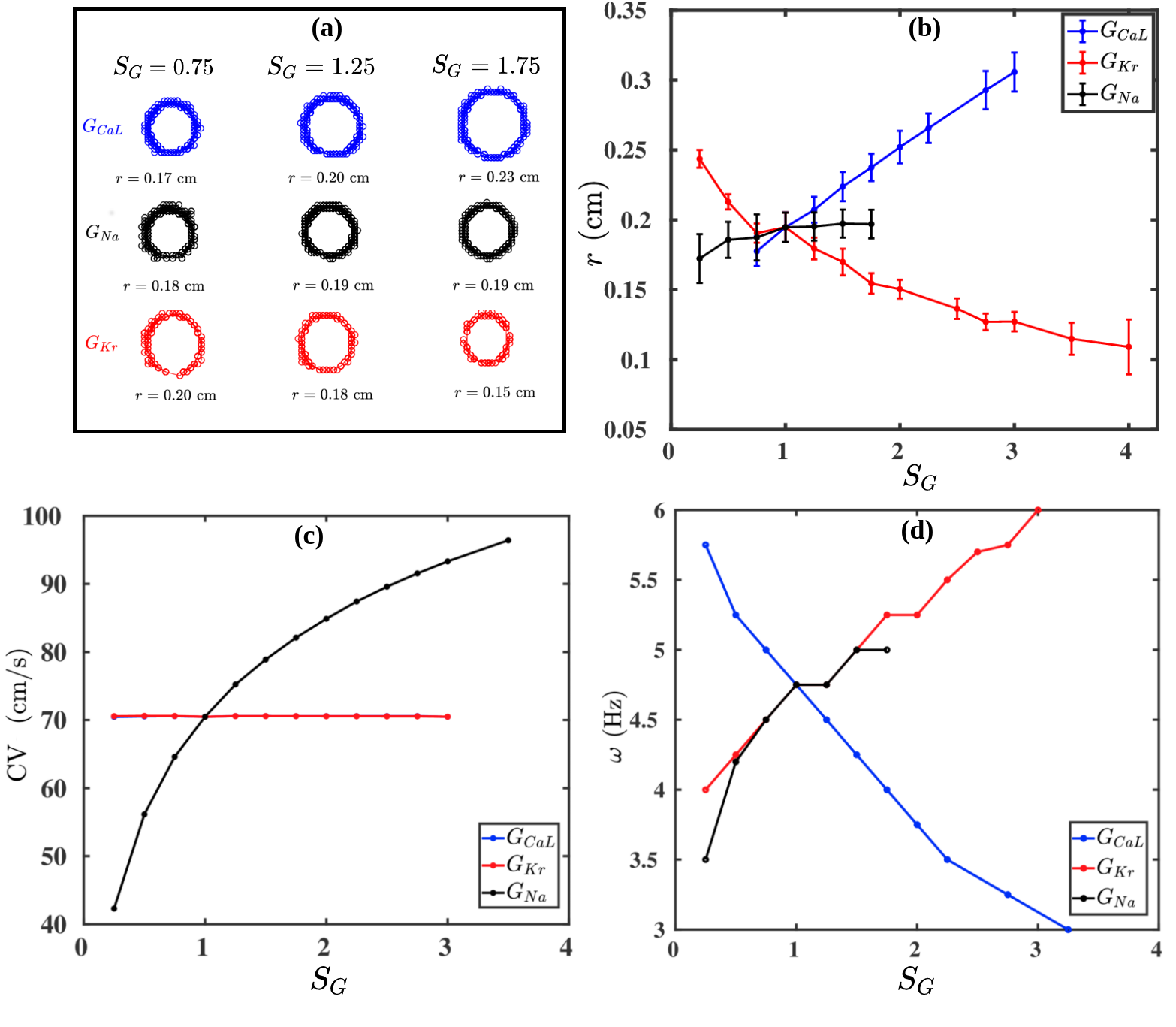}
\caption{(a) Traces of the tip trajectories of a spiral for different values of
conductances of three ion-channels: $G_{CaL}$ (blue), $G_{Na}$ (black),
and $G_{Kr}$ (red); the columns indicate  $S_G$ (Eq.~\ref{eq:nondim}),
which multiplies only the conductance that labels a row (all other
conductances are held at their control values as we move along a row).
(b), (c), and (d): Plots versus $S_G$ of, respectively, $r$, CV, and
$\omega$ (see text), for all these three conductances;
one-standard-deviation error bars are shown for $r$.}  
\label{fig:fig2}
\end{figure*}

We now study how these changes in $\frac{dV}{dt}_{max}$ and the APD affect the
dynamics of a spiral wave. In Fig.~\ref{fig:fig2} (a) we show spiral-tip
trajectories and how the
radius $r$, of the averaged circular trajectory, varies with the three
conductances $G_{CaL}$ (blue), $G_{Na}$ (black), and $G_{Kr}$ (red); the
columns are labelled by the values of  $S_G$ (Eq.~\ref{eq:nondim}), which
multiply only the conductance that labels a row (all other conductances are
held at their control values as we move along a row in  Fig.~\ref{fig:fig2}
(a)).  In Figs.~\ref{fig:fig2} (b), (c), and (d) we give plots versus $S_G$ of,
respectively, $r$, CV, and $\omega$, for all these three conductances.  In
particular, we find that $\omega$ increases if we increase the values of
$G_{Na}$ and $G_{Kr}$; by contrast, $\omega$ decreases as we increase
$G_{CaL}$.  This is consistent with the variation of $r$ and of CV with $S_G$
(Figs.~\ref{fig:fig2} (b) and (c)), for $\omega$ is related to $r$ and CV as in
Eq.~\ref{eqn:omega_exact}.  If we raise the values of $G_{CaL}$ and $G_{Kr}$,
then we find an increase and decrease the spiral core radius $r$, respectively,
whereas $G_{Na}$ has no significant effect on the value of $r$
(Fig.~\ref{fig:fig2} (b)).  Furthermore, Fig.~\ref{fig:fig2} (c) shows that CV
increases with $G_{Na}$, whereas $G_{CaL}$ and $G_{Kr}$ do not affect CV; this
is because only  $G_{Na}$ affects the value of $\frac{dV}{dt}_{max}$
(Fig.~\ref{fig:fig1} (c)), which determines how fast a myocyte is excited and,
therefore, how rapidly a wave of excitation propagates through our
cardiac-tissue model. This result, along with Fig.~\ref{fig:fig1} (b), implies
that the change in the APD is associated with the change in the value of $r$; a
large (small) value of the APD is associated with a large (small) value of $r$;
and conductances such as $G_{Na}$ have no significant effect on the APD because
they do not affect $r$ substantially. We have also checked this correlation
between the APD and $r$ for other conductances (see Fig. S2 in the Supplemental
Material) and have found similar results. In summary, the rise of $\omega$ with
the increase of $G_{Na}$ is primarily because of the increase in CV, and the
decline (rise) of $\omega$, with the increase of $G_{CaL}$ ($G_{Kr}$), can 
be attributed principally because to the increase (decrease) in $r$.         
   
\begin{equation}
\omega \propto \frac{\text{CV}}{2\pi r} \\
\label{eqn:omega_exact}
\end{equation}

\subsection{Effect of the gap-junctional coupling on $\omega$}
\label{subsec:gapjn}

\begin{figure*}
\centering
	\includegraphics[scale=0.25]{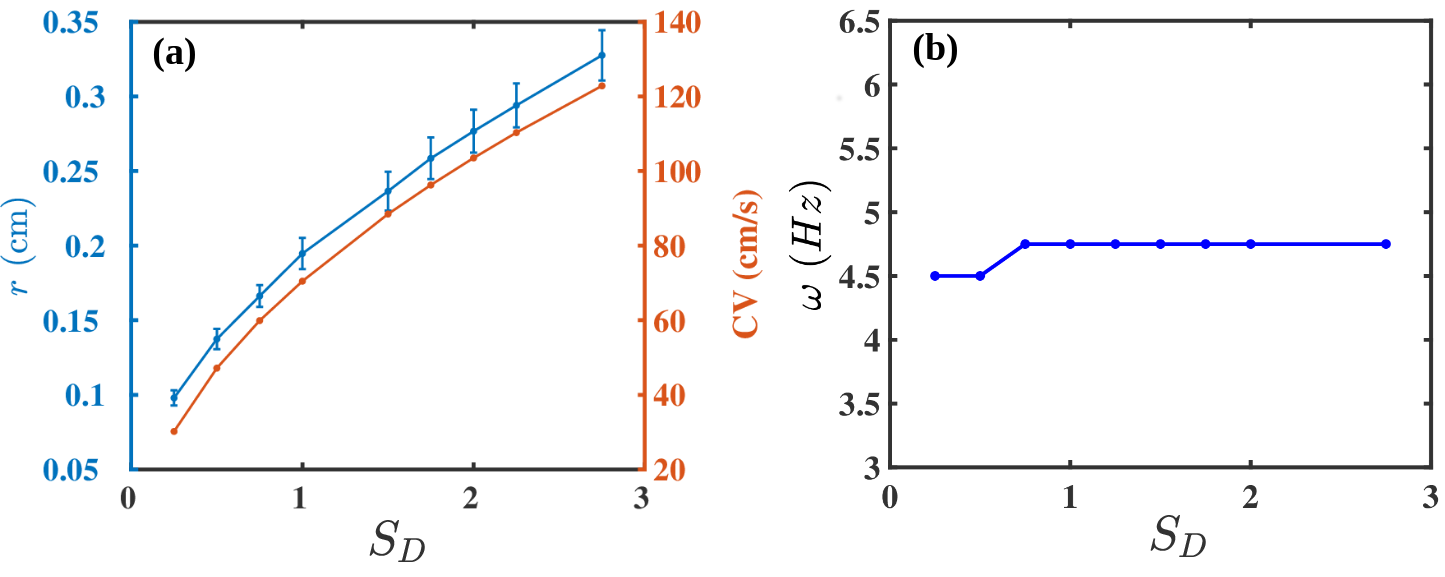} 
	\caption{Plots versus $S_D$ (Eq.~\ref{eq:nondim}) of (a) $r$ (blue curve)
and CV (red curve) and (b) $\omega$.} 
\label{fig:fig3}
\end{figure*}

The strength of the gap-junctional coupling between the cells in cardiac tissue
can change in diseased conditions, e.g., in the wake of a myocardial
infarction~\cite{de1993slow,king2013determinants,mcdowell2011susceptibility}.
It is, therefore, instructive to investigate the role of the diffusive coupling
betwen the cells on spiral-wave dynamics.  To study the effect of $D$ on
$\omega$, we first plot, in Fig.~\ref{fig:fig3} (a), $r$ (blue curve) and CV
(red curve) versus $S_D$, the non-dimensionalised diffusion constant in
Eq.~\ref{eq:nondim}; this shows that both $r$ and CV increase with $S_D$,
because a high diffusive coupling enhances the propagation of waves. The
increase in CV is offset by the increase in $r$, so $\omega$ (see
Eq.~\ref{eqn:omega_exact}) does not depend on $S_D$ significantly, as we show
in Fig.~\ref{fig:fig3} (b).

\begin{figure*}
\centering
	\includegraphics[scale=0.30]{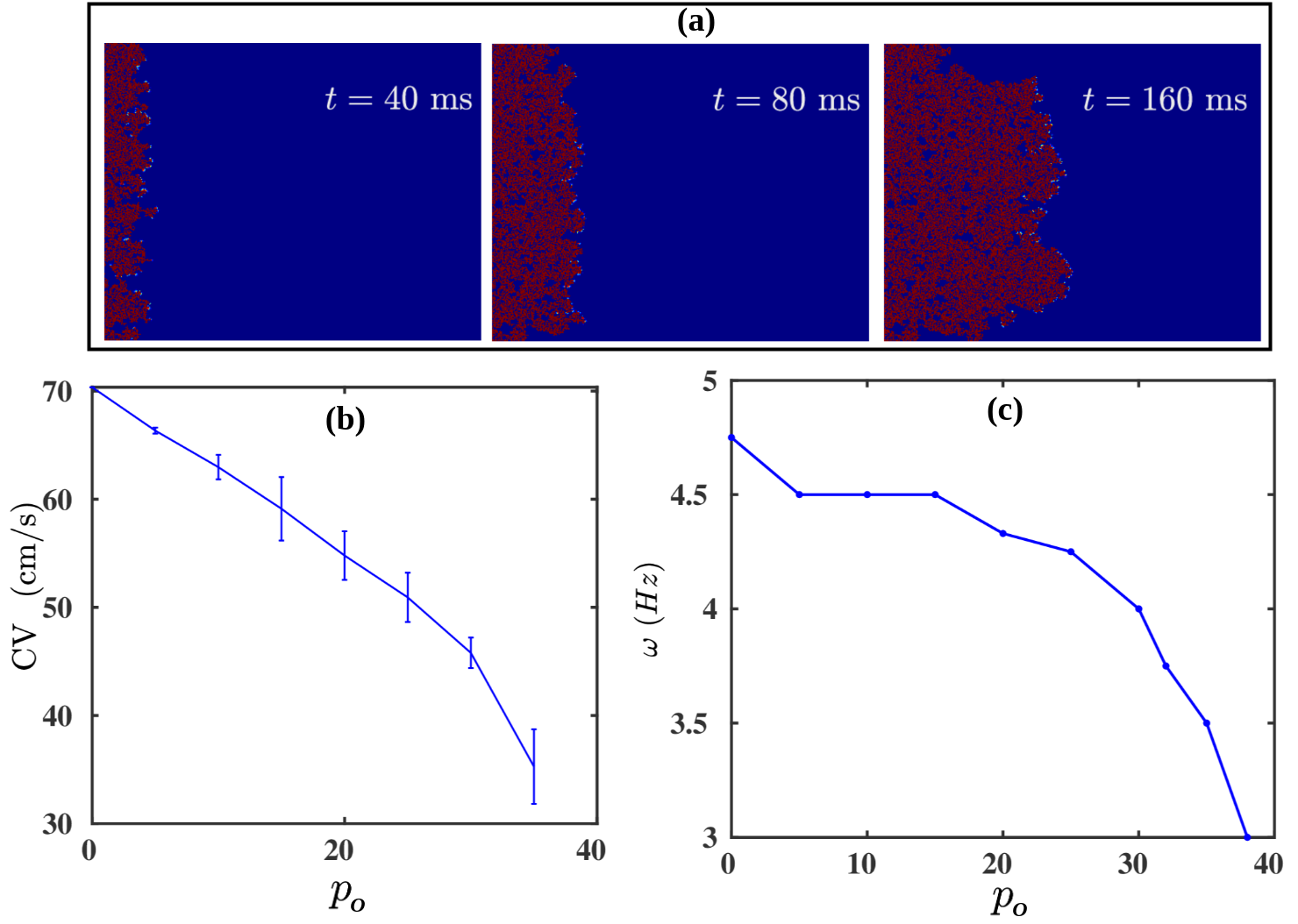}
	\caption{(a) Pseudocolor plots of $V_m$ illustrating the propagation of a 
plane wave through the simulation domain with randomly 
distributed inexcitable obstacles (\textbf{Model I}); 
the obstacle density $p_o = 35 \%$. (b) and (c): Plots versus $p_o$ of the 
plane-wave conduction velocity CV and the spiral-wave frequency $\omega$. CV vanishes after
$p_o \geq 38\%$; i.e., there is \textit{conduction block}.}
\label{fig:fig4}
\end{figure*}

We can also reduce the effective coupling strength between the cells in the
medium by interspersing the medium with inexcitable point obstacles. These
obstacles mimic collagen deposits in fibrotic
tissue~\cite{weber1994collagen,spach1997microfibrosis}. The random distribution
of these obstacles disrupts the propagation of a wave, as we show by the
pseudocolor plots of $V_m$ in Fig.~\ref{fig:fig4} (a); and it reduces the
velocity of the wave~\cite{de1993slow,ten2005wave}. In Fig.~\ref{fig:fig4} (b)
we plot CV versus $p_o$; clearly, CV decreases as the obstacle density $p_o$
increases; and beyond $p_o \simeq 38\%$, we observe \textit{conduction block} 
with CV$= 0$. This
reduction in CV, with the increase of $p_o$, contributes to the decline of
$\omega$ with increasing $p_o$, which we depict by the plot in
Fig.~\ref{fig:fig4} (c). Futhermore, because of the disorder-induced corrugated
wavefront (Fig.~\ref{fig:fig4} (a)), it becomes difficult to track the
spiral-tip trajectory for $p_o > 10\%$; for $p_o < 10\%$, the
value of $r$ remains unaltered (see Fig. S3 in the Supplementary Material).
Nonetheless, the simultaneous decrease of $\omega$ and CV, as we increase
$p_o$, tells us that the change in CV is responsible principally
for the variation of $\omega$.    

%%%%%%%%%%%%%%%%%%%%%%%%%%%%%%%%%%%%%%%%%%%%%%%%%%%%%%%%%%%%%%%%%%%%%%%%%%%%%%%%%%%%%%%%%%%%%%%%

\subsection{Effect of the fibroblast-myocyte coupling on AP properties and $\omega$}
\label{subsec:sec3}

\begin{figure*}
\centering
	\includegraphics[scale=0.25]{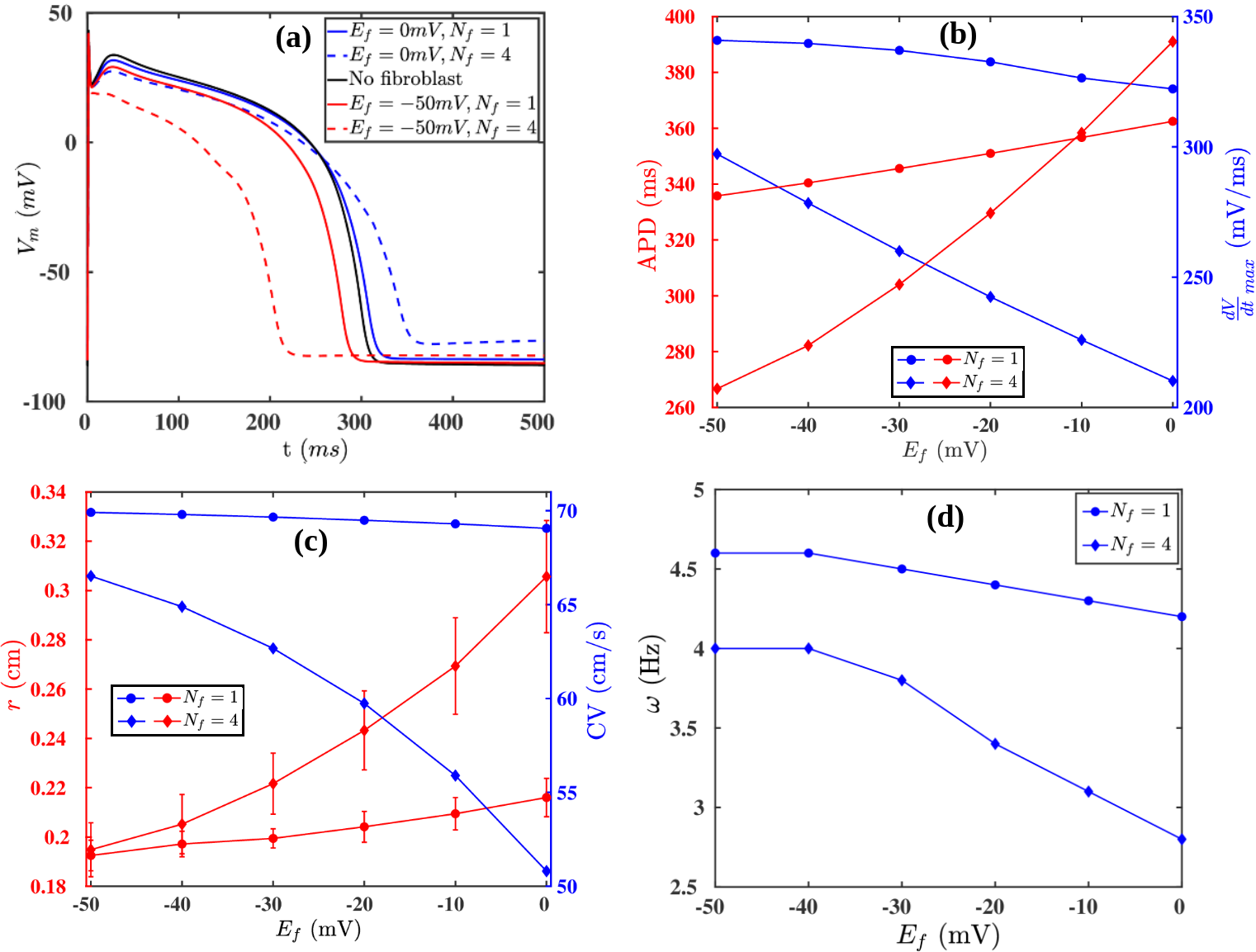}

\caption{(a) APs of an isolated myocyte (black -) and a myocyte coupled to
fibroblasts with various paramters: $E_f$= 0 mV, $N_f=1$ (blue -);
$E_f= 0$ mV, $N_f=4$ (blue --); $E_f= -50$ mV, $N_f=1$ (red -); $E_f=
-50$ mV, $N_f=4$ (red --). (b) The values of the APD and
$\frac{dV}{dt}_{max}$ for different values of $E_f$ and two different
values of $N_f$. (c) The values of $r$ and CV for different values of
$E_f$ and two different values of $N_f$. (d) The varition of $\omega$
with changes in $E_f$ for two different values of $N_f$.}

\label{fig:fig5}
\end{figure*}

Fibroblast cells, which maintain the structural integrity of a heart, are
known to (a) proliferate in diseased conditions~\cite{weber1994collagen,manabe2002gene} 
and (b) form gap-junctional couplings with
myocytes. Such couplings can modulate the
electrophysiological properties, e.g., of the AP, of the mocytes~\cite{jacquemet2007modelling,zlochiver2008electrotonic,maccannell2007mathematical}.
We show in Figs.~\ref{fig:fig5} (a) and (b), how the fibroblast-myocyte coupling
affects the AP morphology, APD, and $\frac{dV}{dt}_{max}$ for different values
of fibroblast resting potential $E_f$ and the number $N_f$ of fibroblasts coupled
to a myocyte in \textbf{Model II}. We see that the APD and $\frac{dV}{dt}_{max}$ 
increase and decrease, respectively, as we increase $E_f$. For a fixed value of $E_f$,
increasing $N_f$ decreases both APD and $\frac{dV}{dt}_{max}$. This is because
fibroblasts act as current sinks when coupled to myocytes. These changes in
the properties of the AP, because of the fibroblast-myocyte coupling, affect 
the wave dynamics at the tissue level. We show in
Fig.~\ref{fig:fig5} (c) that the rise in the APD and the decline in
$\frac{dV}{dt}_{max}$ (see Fig.~\ref{fig:fig5} (b)) increases and decreases the
values of $r$ and CV, respectively, as we increase $E_f$. In Fig.~\ref{fig:fig5}
(d) we show how the combination of these effects on CV and $r$ affect the
variation of $\omega$ with $E_f$ and $N_f$.

\subsection{Drift of spiral waves in domains with an
inhomogeneous distribution of fibroblasts}
\label{subsec:drift}

\begin{figure*}
\centering
\includegraphics[scale=0.35]{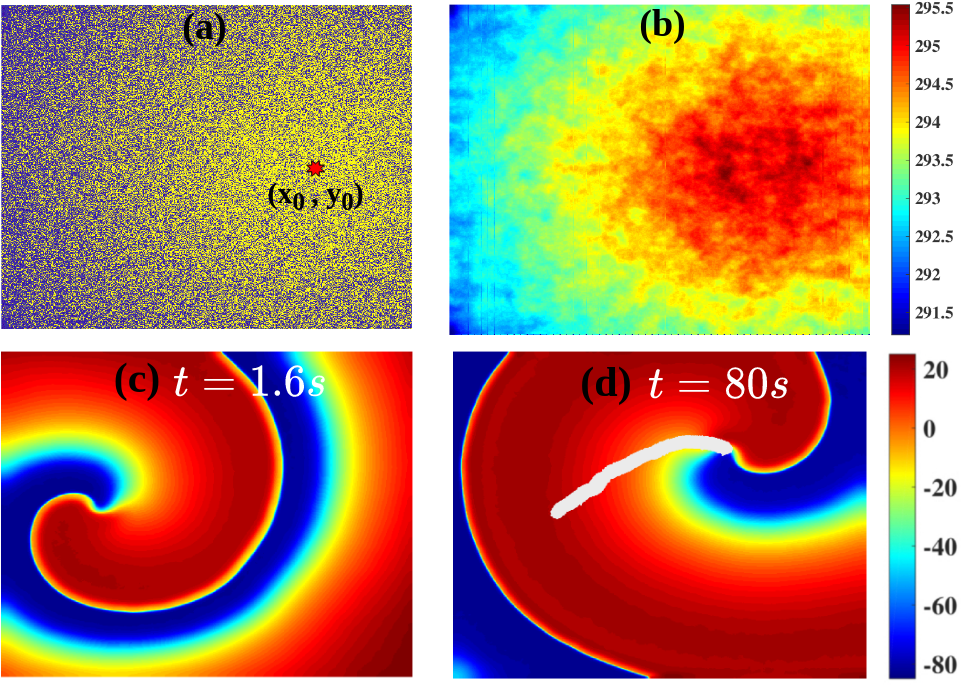}
\caption{(a) The radially decreasing distribution of the
fibroblast density (\textbf{Model-II} Eq.~\ref{eqn:eqn1}) away from a
center, marked by a red octagram; yellow indicates fibroblast-myocyte composites 
and blue denoted myocytes. (b) The distribution of the APD because of the 
gradient in fibroblast density. (c) and (d): pseudocolor plots of $V_m$
showing a spiral wave in the simulation domain: a spiral initiated in the 
small-APD region, proximal to the left boudary, drifts towards the 
large-APD (low-$\omega$) region. The tip trajectory of the spiral is marked by 
the white line.}  
 \label{fig:fig6}
\end{figure*}

Fibrosis is a natural wound-healing process that occurs in the heart after a
patient suffers from a condition such as infarction or heart
attack~\cite{gurtner2008wound,biernacka2011aging,hinderer2019cardiac}, and such
fibrotic tissue can affect the propagation of excitation
waves~\cite{de1993slow,kawara2001activation,ten2005wave,xie2009effects,king2013determinants,morgan2016slow},
which can promote arrhythmias. We now show how a heterogeneous density of
fibroblasts in the medium can affect the dynamics of a spiral wave.
Figure~\ref{fig:fig6} (a) shows the hetereogeneous distribution of fibroblasts
in the medium; here, yellow indicates fibroblast-myocyte composites and blue
indicates myocytes. The density of fibroblasts decreases radially outwards from
the centre that is marked by a red octagram in Fig.~\ref{fig:fig6} (a)
(Sec.~\ref{sec:materials} Eq.~\ref{eqn:eqn1}).  Figure~\ref{fig:fig6} (b) shows
the spatial variation of the APD in the medium because of the heterogeneous
fibroblast density. Figures~\ref{fig:fig6} (c) and (d) show the spatiotemporal
evolution of a spiral in this case. It shows that a spiral, initiated at the
left side of the domain in the region with a low density of fibroblasts, drifts
towards the region with a high density of fibroblasts; and the spiral remains
anchored to the central region, where the fibroblast density is maximum. The
trajectory of the spiral tip is shown in white in Fig.~\ref{fig:fig6} (d) (see
also the video M1 in the Supplemental Material). This drifting of a spiral
towards the region with a high density of fibroblasts is associated with the
tendency of the spiral wave to drift towards a region high values of the
APD~\cite{rudenko1983drift,krinsky1996dense,qu2005effects,sridhar2010anomalous,berenfeld2016major}.
Such anchoring of a spiral wave to a region with a high density of fibrosis has
been seen in experiments on real
hearts~\cite{fast1990drift,davidenko1992stationary,jalife1996drifting,ten2003reentry,roney2016modelling}.
Our study illustrates how a region with a high density of fibroblasts can
behave like an attractor and an anchoring point for spiral waves in fibrotic
tissue. Such drifting of a spiral wave, in a medium with heterogeneity, has also
been reported in other studies in contexts other than fibrosis~\cite{rudenko1983drift,krinsky1996dense,qu2005effects,sridhar2010anomalous,berenfeld2016major}.        
      
\section{Discussion}
\label{sec:discussion}

We have used \textit{in silico} simulations of detailed mathematical models for
cardiac tissue to examine the effects of various electrophysiological
paramaters of a cardiac cell and cardiac tissue on the AP properties and on
electrical-wave dynamics. Our work is of relevance to such waves in real
hearts, which are intrinsically heterogeneous along the
transmural~\cite{wolk1999functional,antzelevitch2001electrical,mccrossan2004transmural}
and the apico-basal~\cite{burton2001dispersion,szentadrassy2005apico}
directions. Moreover, heterogeneities can be also be induced in the heart
because of
diseases~\cite{viswanathan2000cellular,burton2001dispersion,schmidt2007infarct,antzelevitch2007heterogeneity}.
In this context, we have shown how changes in various ion-channel conductances
of a myocyte or the fibroblast-myocyte coupling can modulate the AP of a
myocyte. We have then checked how these changes affect the spiral-wave
frequency $\omega$. We find that an increase (decrease) in
$\frac{dV}{dt}_{max}$ or decrease (increase) in the APD increases (decreases)
$\omega$: large values of $\frac{dV}{dt}_{max}$ increase CV; and a low APD is
associated with low values of the mean spiral-tip-trajectory radius $r$; these
are related to $\omega$ through Eq.~\ref{eqn:omega_exact}. Our study has
provided a natural understanding of how changes in the AP, at the
single-myocyte level, can be related to changes in $\omega$ at the
cardiac-tissue level. Moreover, we have investigated how changes in the
gap-junctional coupling between the cells and $S_D$ affect $\omega$.  We have
also reduced the effective coupling between the cells by interspersing the
medium with inexcitable obstacles; $\omega$ changes with the density of the
obstacles. It is of interest to investigate such effects on $\omega$, because
they provide insights into spiral-wave dynamics in excitable media with
heterogeneities~\cite{antzelevitch2007heterogeneity}. We illustrate this in
detail in Fig.~\ref{fig:fig6} for a simulation domain with a heterogeneous
distribution of fibroblast; here, we demonstrate the drift of a spiral wave
towards the region with a high density of fibroblasts; such a drift has been
seen in real
hearts~\cite{fast1990drift,davidenko1992stationary,jalife1996drifting,roney2016modelling}.

\textcolor{black}{We have explored the validity of
the frequency relation~\ref{eqn:omega_exact} (Ref.~\cite{qu2014nonlinear}) 
for a wide range of electrophysiological parameters in the models that 
we use. We show in Fig.S7 in the Supplemental Material
Ref.~\cite{supmat_EP} that our measurements of $\omega$ and $\frac{\text{CV}}{r}$
are consistent with a linear relation (see the fit that is indicated by a 
black line); at very low values of CV, e.g., near conduction block in 
Model I which accounts for fibrosis-induced disorder, this 
linear relation breaks down. The randomness in these models
introduces error in the determination of $r$ of the spiral wave,
especially for large randomness; e.g., as we increase $p_f$, we observe, 
in Fig.S6 (Supplemental Material~\cite{supmat_EP}) that the tip
trajectory of the spiral wave becomes very noisy. 
Note also that the CV of a plane wave is distinct from CV$_{tip}$ the velocity of the tip of the spiral wave 
as it goes around its trajectory (on average a circle with radius $r$); clearly,
$\omega_{tip} = \text{CV}_{tip}/(2 \pi r)$ (see Table.S1 in the Supplemental Material~\cite{supmat_EP}).}

Some earlier studies have investigated the 
properties of spiral waves in two-variable mathematical
models for cardiac~\cite{mikhailov1983rotating,winfree1991varieties,mikhailov1994complex,hakim1999theory,margerit2002cookbook,zykov2009kinematics,lober2013analytical}.
However, such studies have been conducted in the weak- or strong-excitability
limits; real cardiac tissue exhibits various degrees of excitability depending
on different electrophysiological parameters. Our study, which employs
electrophysiologically detailed mathematical models for cardiac tissue, has
allowed us to study spiral-wave dynamics with greater realism than is possible
with two-variable models for cardiac tissue. The drifting of a spiral wave
towards regions with a large APD has been reported in contexts other than
fibrosis~\cite{rudenko1983drift,krinsky1996dense,qu2005effects,sridhar2010anomalous,berenfeld2016major}.
Moreover, anomalous drift of a spiral towards a region with a small APD, which
has been observed in generic models~\cite{sridhar2010anomalous}, is not seen in our study; and it is yet to
be reported in any in any of the electrophysiologically-detailed mathematical models for the cardiac tissue. 
It is also obeserved in the two-variable models that the radius of the spiral tip
trajectory is very large, in the weakly excitable limit, compared to what is
observed in the strongly excitable
limit~\cite{barkley1994euclidean,hakim1999theory}. In case of our realistic models, if
we consider two parameters that control excitability, e.g. , $G_{Na}$ and $D$,
then we observe that $r$ does not increase with a decrease in the value of 
$G_{Na}$ (see Fig.~\ref{fig:fig2}); but we observe an increase in $r$, as we
increase the value of $D$ (see Fig. S5 of the Supplemental 
Material~\cite{supmat_EP}. Hence our systematic study, which uses a detailed
human-ventricular-tissue mathematical model, provides an important point of
reference for future \textit{in silico} and experimental studies of such spiral
waves in cardiac tissue.   

We end our discussion with some limitations in our study. We have used a
monodomain model for cardiac tissue. Bidomain models of cardiac tissue account
for the extracellular matrix.  However, monodomain models have been proved to
be good approximations of cardiac tissue for wave
propagation~\cite{potse2006comparison} for the types of excitations we
consider. Furthermore, our tissue model does not incorporate the effects of
mechanical deformations, stretch-activated channels, and stress-dependent
diffusion
tensors~\cite{zeng2000stretch,kamkin2000stretch,thompson2011mechanical}.  Such
deformations can affect the dynamics of spiral waves~\cite{panfilov2007drift}
and the drift of spirals in a heterogeneous medium; we defer an investigation
of the interplay between deformation and drift for future work.

\section*{References}
\bibliography{references}

\end{document}

\title{Supplementary Material: An {\it in silico} study of electrophysiological parameters 
that affect the spiral-wave frequency in cardiac tissue}
\author{Mahesh Kumar Mulimani}
\email{maheshk@iisc.ac.in ;}
\affiliation{Centre for Condensed Matter Theory, Department of Physics, Indian Institute of Science, Bangalore 560012, India.}
\author{Soling Zimik}
\email{solyzk@gmail.com ;}
\affiliation{Centre for Condensed Matter Theory, Department of Physics, Indian Institute of Science, Bangalore 560012, India.}
\author{Rahul Pandit}
\email{rahul@iisc.ac.in}
\altaffiliation[\\]{also at Jawaharlal Nehru Centre For
Advanced Scientific Research, Jakkur, Bangalore, India}
\affiliation{Centre for Condensed Matter Theory, Department of Physics, Indian Institute of Science, Bangalore 560012, India.}

\pacs{87.19.Xx, 87.15.Aa }

\maketitle

\begin{figure*}
	\centering
        \includegraphics[scale=0.35]{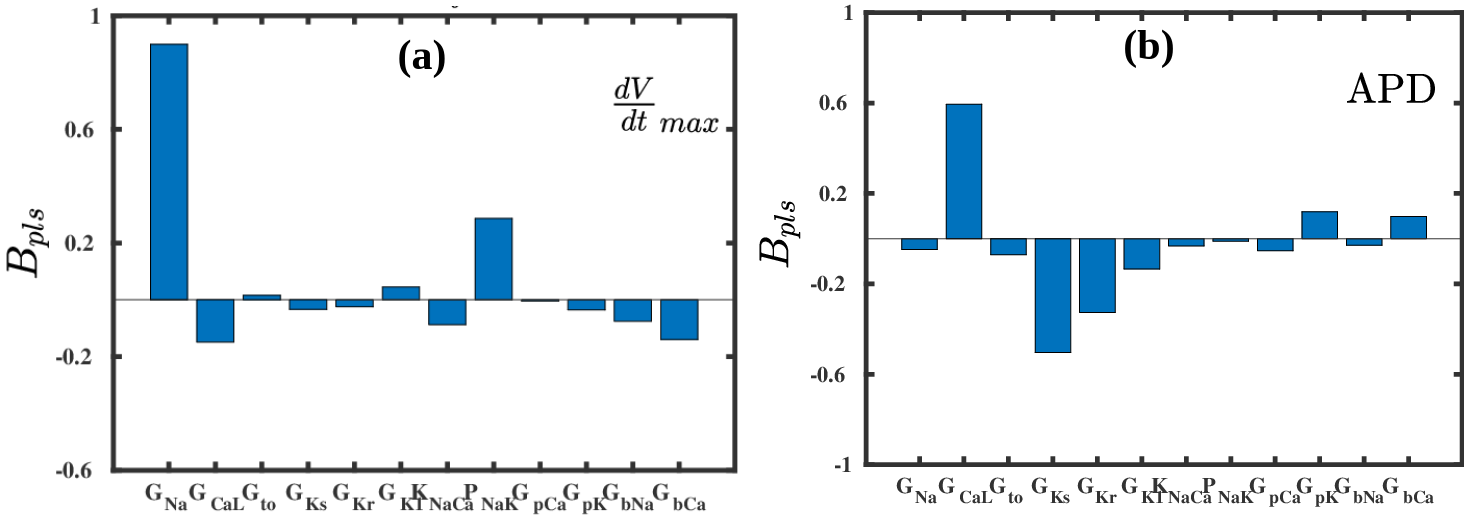}
\caption{$B_{pls}$, shown on
the vertical axes, is the regression coefficient that gives the
extent to which the parameter, shown on the horizontal axis, controls
(a) $\frac{dV}{dt}_{max})$ and (b) the APD. Positive (negative) value of $B_{pls}$
imply that an increase in the parameter increases (decreases)
the value of (a) $\frac{dV}{dt}_{max})$ or (b) the APD. For the details of sensitivity 
analysis see Ref.~\cite{mulimani2020comparisons}.}		
\label{fig:figs1}
    \end{figure*}

\begin{figure*}
\centering
\includegraphics[scale=0.18]{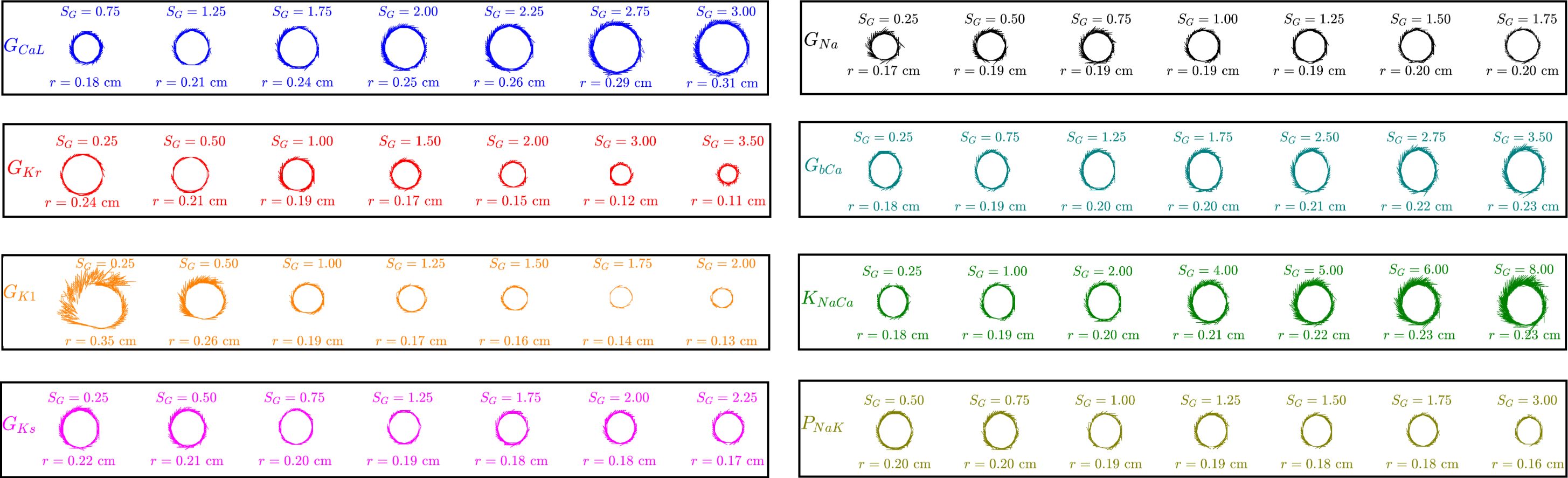}
\caption{Spiral-tip trajectories illustrating how the
radius $r$, of the averaged circular trajectory, varies with the
different conductances and ion-pump parameters in the TP06 model; the columns are
labelled by the values of  $S_G$ (Eq.6 in the
main paper), which multiply only the conductance or ion-pump
parameter that labels a row (all other conductances are held at
their control values as we move along a row in
Fig.~\ref{fig:figs11}); $G_{to}, G_{pK} \text{and} \ G_{bNa}$, do not affect $r$, 
CV, and $\omega$ significantly.}		
\label{fig:figs11}
    \end{figure*}

\begin{figure*}
	\centering
        \includegraphics[scale=0.32]{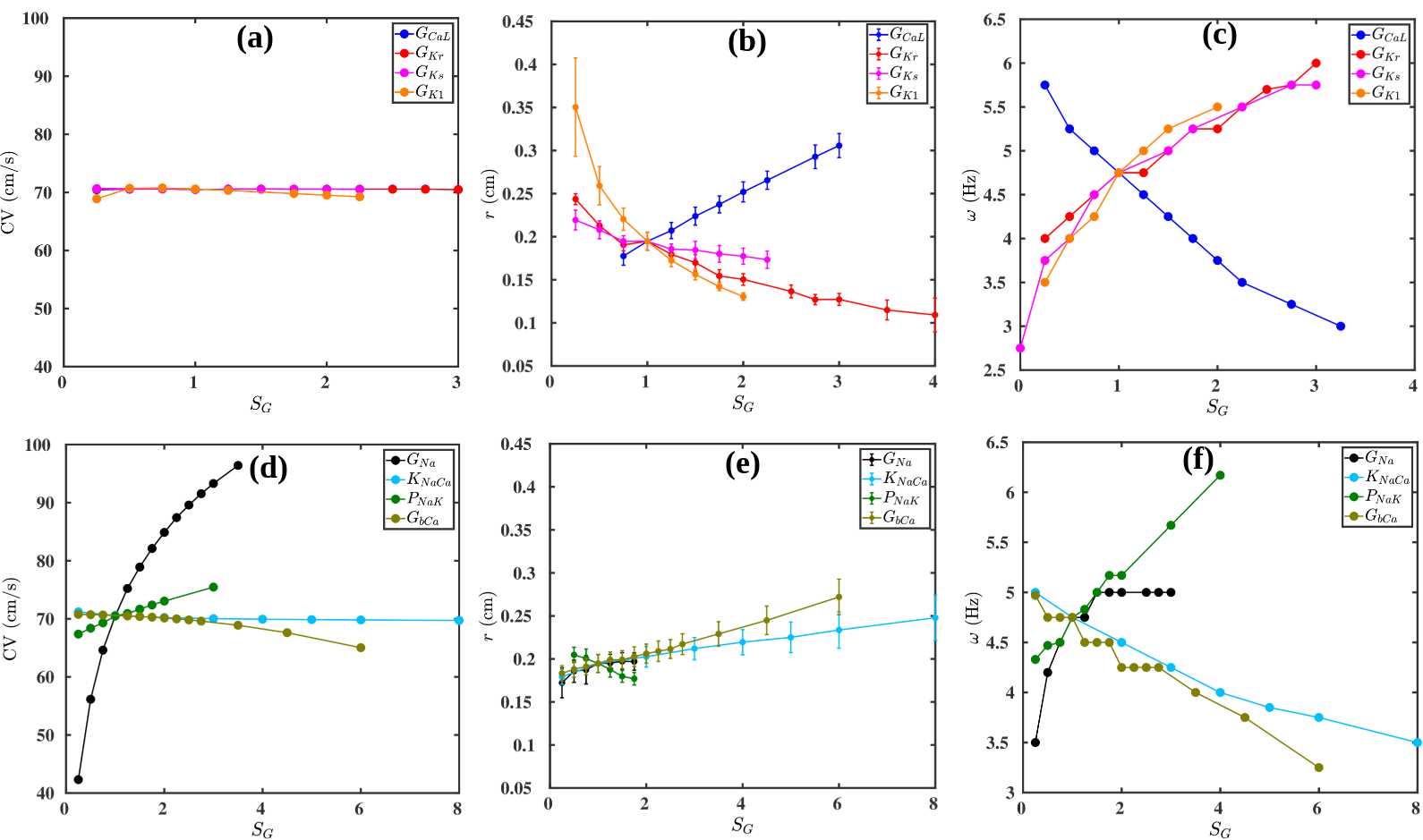}
\caption{(a)-(f) Plots versus $S_G$ of $r$, CV,
and $\omega$ (see the main paper), for various conductances and
ion-pump parameters in the TP06 model.}		
\label{fig:figs13}
    \end{figure*}

\begin{figure*}[!htb]
	\centering
       \includegraphics[scale=0.40]{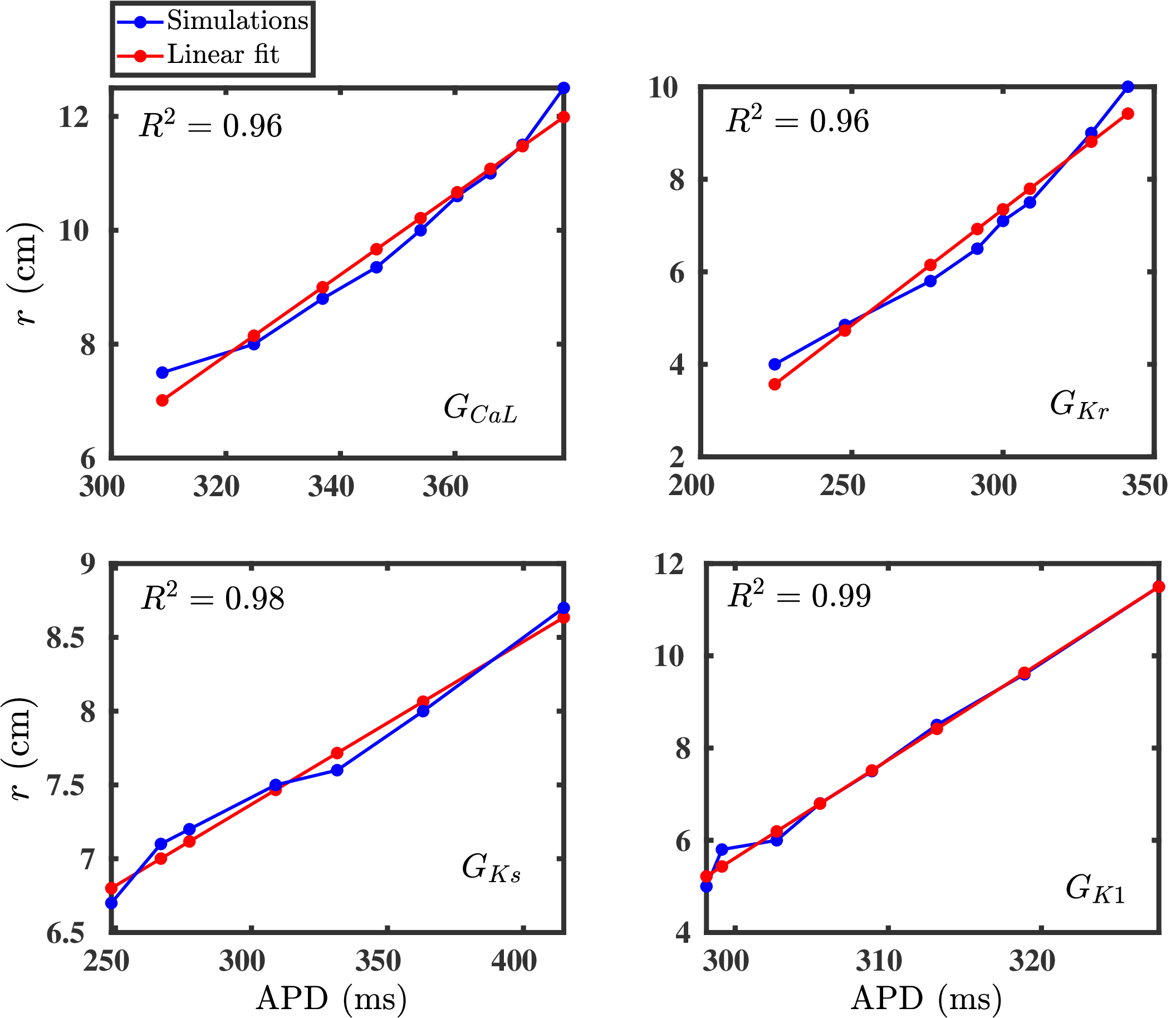}
	\caption{(a)-(d) Plots showing the dependence of $r$ on the
APD, which is tuned by changing the conductances (a) $G_{CaL}$, (b)
$G_{Kr}$, (c) $G_{Ks}$ and $G_{K1}$, respectively. 
$R^2$ in the plots show the goodness of the fit.}
\label{fig:figs2}
\end{figure*}

\begin{figure*}[!htb]
	\centering
        \includegraphics[scale=0.28]{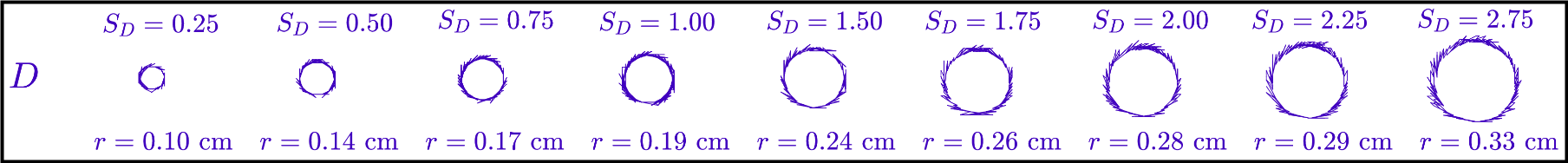}
	\caption{Spiral-tip trajectories illustrating how the
radius $r$, of the averaged circular trajectory, varies with $S_D$, the
non-dimensionalised diffusion coefficient (Eq.6
in the main paper); all conductances and ion-pump parameters
are held at their control values.}
\label{fig:figs14}
\end{figure*}

\begin{figure*}
       \includegraphics[scale=0.40]{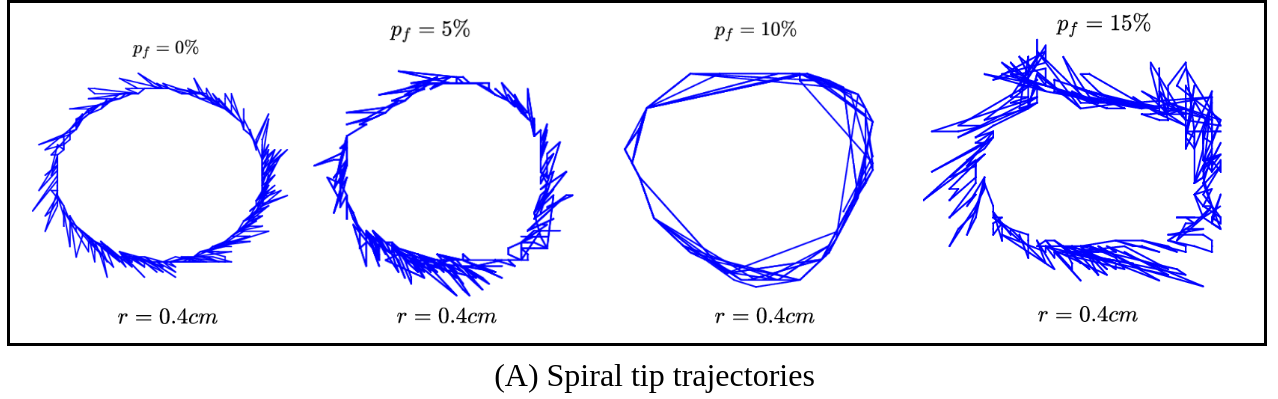}
\caption{Tip trajectories of spiral waves in a
tissue with randomly dispersed inexcitable obstacles (\textbf{Model I} in
the main paper) for four different values of $p_o$.}
\label{fig:figs3}
\end{figure*}

\begin{figure*}
	\centering
        \includegraphics[scale=0.35]{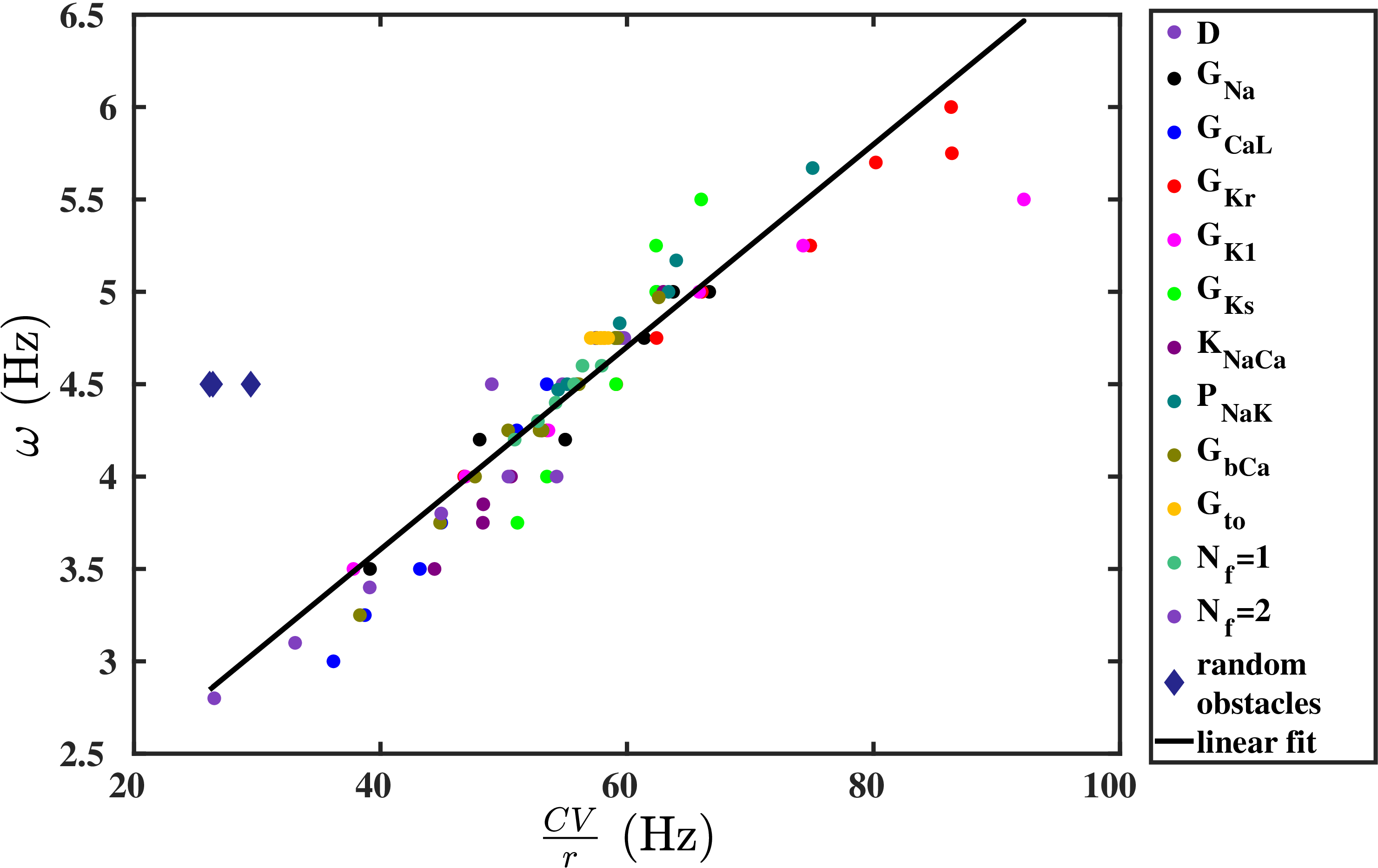}
\caption{The linear fit (black line) to data points from our measurements of $\omega$ and 
$\frac{\text{CV}}{2\pi r}$ for all the electrophysiological parameters 
we use; the diamonds show data points from two of our studies 
with randomly distributed inexcitable obstacles. The quality of the fit, for therange of values shown is $R^2 = 88.9 \%$, if we leave out the points indicated
	by diamonds.}
\label{fig:figs4}
    \end{figure*}

We give below the following supplementary information, which is mentioned in the 
main paper:

\begin{enumerate}

\item We show in Figs.~\ref{fig:figs1} (a) and (b) the contributions of various
ion-channel conductances and ion-pump parameters to the upstroke-velocity
(a) $\frac{dV}{dt}_{max}$ and (b) the APD, respectively. We then use parameter-sensitivity 
analysis, specifically the Principal-Least-Squares
regression analysis (PLS), which is  described in detail in
Ref.~\cite{mulimani2020comparisons}. We generate these data by pacing the cell
with a pacing-cycle-length PCL$=1$s; and we record the last, steady-state AP and
$\frac{dV}{dt}_{max})$. $B_{pls}$, shown on
the vertical axes, is the regression coefficient that gives the
extent to which the parameter, shown on the horizontal axis, controls
(a) $\frac{dV}{dt}_{max})$ and (b) the APD. Positive (negative) value of $B_{pls}$
imply that an increase in the parameter increases (decreases)
the value of (a) $\frac{dV}{dt}_{max})$ and (b) the APD. We observe that
$G_{Na}$ controls $\frac{dV}{dt}_{max})$ most; this is
followed by the $Na-K$ ion-exchanger pump ($P_{NaK}$); the
other conductances and ion-pump parameters do not affect 
$\frac{dV}{dt}_{max})$ siginificantly. 

\item In Fig.~\ref{fig:figs11} we show spiral-tip trajectories and how the
radius $r$, of the averaged circular trajectory, varies with the
different conductances and ion-pump parameters in the TP06 model; the columns are
labelled by the values of  $S_G$ (Eq.6 in the
main paper), which multiply only the conductance or ion-pump
parameter that labels a row (all other conductances are held at
their control values as we move along a row in
Fig.~\ref{fig:figs11}).
	
\item In Figs.~\ref{fig:figs13} (a)-(f) we give plots versus $S_G$ of $r$, CV,
and $\omega$ (see the main paper), for various conductances and
ion-pump parameters in the TP06 model. 

\item In Figs.~\ref{fig:figs2} (a)-(d) we show the dependence of $r$ on the
APD, which is tuned by changing the conductances (a) $G_{CaL}$, (b)
$G_{Kr}$, (c) $G_{Ks}$ and $G_{K1}$, respectively.

\item In Fig.~\ref{fig:figs14} we show spiral-tip trajectories and how the
radius $r$, of the averaged circular trajectory, varies with $S_D$, the
non-dimensionalised diffusion coefficient (Eq.6
in the main paper); all conductances and ion-pump parameters
are held at their control values.

\item In Fig.~\ref{fig:figs3} we show tip trajectories of spiral waves in a
tissue with randomly dispersed inexcitable obstacles (\textbf{Model I} in
the main paper) for four different values of $p_o$.

\item \textcolor{black}{In Fig.~\ref{fig:figs4} we show the linear fit (black line) to data points from our measurements of $\omega$ and 
$\frac{\text{CV}}{2\pi r}$ for all the electrophysiological parameters 
we use; the diamonds show data points from two of our studies
with randomly distributed inexcitable obstacles. The quality of the fit, for therange of values shown is $R^2 = 88.9\%$, if we leave out the points indicated 
by diamonds.}

\item \textcolor{black}{In Table.S1, we show for different electrophysiological
parameters the observed frequency of the spiral tip $\omega_{tip}$ and
the frequency of the spiral wave calculated by using Fourier
analysis of the time series data $\omega \equiv
\omega_{Fourier}$ (see the main paper). We calculate the
$\omega_{tip}$ by noting down the time duration where the
spiral tip completes one complete rotation.
\begin{eqnarray}
\omega_{tip}= \frac{\text{CV}_{tip}}{(2\pi r)} \\
\text{CV}_{tip} = \frac{2\pi r} {T} \\
\omega_{tip}= \frac{1}{T}
\label{eq:omega_tip}
\end{eqnarray}
$T$ is the period of the spiral tip. 
}

\item \textbf{Movie-M1}: A movie from our \textit{in silico}
study showing the drift of a 
spiral wave (via pseudocolor plots of the transmembrane potential $V_m$)
in the presence of gradient in the fibroblast density; the black star in
the movie indicates the position of the highest fibroblast
density (see text in the main paper). 

\end{enumerate}

%%%%%%%%%%%%%%%%%%%%%%%%%%%%%%%%%%%%%%%%%%%%%%%%%%%%%%%%%%%%%%%%%%%%%%%%%%%%%%%%%%%%%%%%%%%%%%%%%%

\begin{table}
\begin{center}
    \begin{tabular}{ | l | l | l | l |}
    \hline 
    Electrophysiological & Time period  & $\omega_{tip} = \frac{1}{T}$ & $\omega_{Fourier}$ \\ 
     parameter    &  of spiral tip (T) in (ms)  & (Hz)  & (Hz)  \\ \hline
     \hline
    \ $G_{Na_{max}} \times 0.75$ & \  220 \ &  \  4.54  \ & \ 4.5  \\ \hline 
    \ $G_{CaL_{max}} \times 0.75$ & \ 200 \  & \  5 \ & \ 5   \\ \hline
    \ $G_{Kr_{max}} \times 2.75$ & \ 210 \ & \  5.77 \ & \ 5.75  \\ \hline 
    \ $D \times 2.75$ & \ 210 & \ 4.76 \ & \ 4.75     \\ \hline
    \ $p_f = 5\%$ & \ 220 & \ 4.54 \ & \ 4.5    \\ \hline
    \ $p_f = 10\%$ & \ 225 & \ 4.44 \ & \ 4.5   \\ \hline
    \ $p_f = 15\%$ & \ 240 & \ 4.17 \ & \ 4.5   \\ \hline 
    \ $p_f = 30\%$ & \ 260 & \ 3.85 \ & \ 4.5    \\ \hline
    \end{tabular}
	\hspace{7cm}\caption{The table shows the observed frequency of the spiral tip $\omega_{tip}$ (see Eq.~\ref{eq:omega_tip}) and the frequency calculated from the Fourier analysis $\omega_{Fourier}$ (see text) for a few representative electrophysiological parameters. }
\end{center}
\label{Table:Table2}
\end{table}     

%%%%%%%%%%%%%%%%%%%%%%%%%%%%%%%%%%%%%%%%%%%%%%%%%%%%%%%%%%%%